\documentclass[a4paper,12pt,twoside]{article}
\usepackage[english]{babel}
\usepackage{rotating}
\usepackage{epsfig}
\usepackage[centertags]{amsmath}
\usepackage{array}
\usepackage{multirow}
\usepackage{supertabular}
\usepackage{dcolumn}
\usepackage{cite}
\usepackage{xspace}
\def\journal#1#2#3#4{{#1}{\bf #2},{ #4} (#3)}
\def\JP{{ J. Phys.\ }}

\def\EPJ{{ Eur.\ Phys.\ J.\ }}
\def\NIM{{ Nucl.\ Instr.\ Meth.\ }}
\def\NP{{ Nucl.\ Phys.\ }}
\def\PL{{ Phys.\ Lett.\ }}

\def\PR{{ Phys.\ Rev.\ }}
\def\PRep{{ Phys.\ Rep.\ }}
\def\ZP{{ Z. Phys.\ }}

\def\IJMP{{ Int.\ J.\ Mod.\ Phys.\ } }

\def\CPC{{ Comp.\ Phys.\ Comm.\ }}
\chardef\usc=95
\chardef\til=126
\catcode`\@=11 
\DeclareRobustCommand\xdotspace{\futurelet\@let@token\@xdotspace}
\def\@xdotspace{%
  \ifx\@let@token.\else
  \ifx\@let@token\bgroup.\else
  \ifx\@let@token\egroup.\else
  \ifx\@let@token\/.\else
  \ifx\@let@token\ .\else
  \ifx\@let@token~.\else
  \ifx\@let@token!.\else
  \ifx\@let@token,.\else
  \ifx\@let@token:.\else
  \ifx\@let@token;.\else
  \ifx\@let@token?.\else
  \ifx\@let@token/.\else
  \ifx\@let@token'.\else
  \ifx\@let@token).\else
  \ifx\@let@token-.\else
  \ifx\@let@token\@xobeysp.\else
  \ifx\@let@token\space.\else
  \ifx\@let@token\@sptoken.\else
   .\space
   \fi\fi\fi\fi\fi\fi\fi\fi\fi\fi\fi\fi\fi\fi\fi\fi\fi\fi}
\catcode`\@=12 

\newcommand{\stru}[2]{%
   \relax\ifmmode\hbox{\vrule height#1 depth#2 width0pt}%
   \else\vrule height#1 depth#2 width0pt\fi}

\newcommand{\Ronum}[1]{\uppercase\expandafter{\romannumeral#1}}
\newcommand{\ronum}[1]{\expandafter{\romannumeral#1}}
\DeclareRobustCommand{\LaTeXZ}{%
  \LaTeX\kern-.05em4\kern-.1em
  {\raisebox{-0.2ex}{$\scriptstyle\text{ZEUS}$}}\xspace}

\DeclareMathAlphabet{\mathbf}{OT1}{cmr}{bx}{sl}
\newcommand{\eVdist}{\kern-0.04em}

\newcommand{\mev}{{\,\text{Me}\eVdist\text{V\/}}}
\newcommand{\gev}{{\,\text{Ge}\eVdist\text{V\/}}}

\newcommand{\slashfrac}[2]{%
  \raisebox{0.5ex}{\ensuremath #1}\kern-0.12em/\kern-0.08em
  \raisebox{-.8ex}{\ensuremath #2}}

\newcommand{\sqr}[3]{%
    {\vcenter{\hrule height.#3ex\hbox{\vrule width.#2ex height#1ex
     \kern#1ex\vrule width.#3ex}\hrule height.#2ex}}}

\catcode`\@=11 
\newcommand{\parenbar}{\mathpalette\p@renb@r}
\def\p@renb@r#1#2{\vbox{%
  \ifx#1\scriptscriptstyle \dimen@.7em\dimen@ii.2em\else
  \ifx#1\scriptstyle \dimen@.8em\dimen@ii.25em\else
  \dimen@1em\dimen@ii.4em\fi\fi \offinterlineskip
  \ialign{\hfill##\hfill\cr
    \vbox{\hrule width\dimen@ii}\cr
    \noalign{\vskip-.3ex}%
    \hbox to\dimen@{$\mathchar300\hfil\mathchar301$}\cr
    \noalign{\vskip-.3ex}%
    $#1#2$\cr}}}
\catcode`\@=12 

\newcommand{\IP}{{\rm I$\kern-0.01667em$P}\xspace}

\mathchardef\qsm=63
\mathchardef\pls=43
\mathchardef\mns=512
\mathchardef\plm=518
\mathchardef\eql=61
\mathchardef\smallleft=300
\mathchardef\smallright=301
\mathchardef\les=316
\mathchardef\gre=318
\mathchardef\leq=532
\mathchardef\grq=533

\catcode`\@=11 
\newcounter{pict@width}
\newcounter{pict@height}
\newlength{\pict@scale}
\newcommand{\psfigadd}[4]{%
\setcounter{pict@width}{1*\ratio{#2+\pict@scale/2}{\pict@scale}}
\setcounter{pict@height}{1*\ratio{#3+\pict@scale/2}{\pict@scale}}
\setlength{\unitlength}{\pict@scale}
\hbox to #2{\hspace{-\fill}\begin{picture}(\thepict@width,\thepict@height)
\put(0,0){\psfig{figure=#1,width=#2,height=#3,clip=}}
\SetScale{0.283466457}
\SetWidth{1.763889}
{#4}
\end{picture}}
}
\newcounter{pict@widthfst}
\newcounter{pict@widthscd}
\newcounter{pict@widthtot}
\newcommand{\psfigaddtwo}[7]{%
\setcounter{pict@widthfst}{1*\ratio{#2+\pict@scale/2}{\pict@scale}}
\setcounter{pict@widthscd}{1*\ratio{#2+#4+\pict@scale/2}{\pict@scale}}
\setcounter{pict@widthtot}{1*\ratio{#2+#4+#6+\pict@scale/2}{\pict@scale}}
\setcounter{pict@height}{1*\ratio{#3+\pict@scale/2}{\pict@scale}}
\setlength{\unitlength}{\pict@scale}
\hbox{\hspace{-\fill}\begin{picture}(\thepict@widthtot,\thepict@height)
\put(0,0){\psfig{figure=#1,width=#2,height=#3,clip=}}
\put(\thepict@widthscd,0){\psfig{figure=#5,width=#6,height=#3,clip=}}
\SetScale{0.283466457}
\SetWidth{1.763889}
{#7}
\end{picture}}
}
\newcommand{\psfigror}[4]{%
\setcounter{pict@width}{1*\ratio{#2+\pict@scale/2}{\pict@scale}}
\setcounter{pict@height}{1*\ratio{#3+\pict@scale/2}{\pict@scale}}
\setlength{\unitlength}{\pict@scale}
\hbox{\begin{picture}(\thepict@width,\thepict@height)
\put(0,\thepict@height){\psfig{figure=#1,width=#3,height=#2,clip=,angle=270}}
\SetScale{0.283466457}
\SetWidth{1.763889}
{#4}
\end{picture}}
}
\newcommand{\psfigrol}[4]{%
\setcounter{pict@width}{1*\ratio{#2+\pict@scale/2}{\pict@scale}}
\setcounter{pict@height}{1*\ratio{#3+\pict@scale/2}{\pict@scale}}
\setlength{\unitlength}{\pict@scale}
\hbox{\begin{picture}(\thepict@width,\thepict@height)
\put(0,0){\psfig{figure=#1,width=#3,height=#2,clip=,angle=90}}
\SetScale{0.283466457}
\SetWidth{1.763889}
{#4}
\end{picture}}
}
\catcode`\@=12 
\newlength\listtextwidth

\catcode`\@=11 
\newlength{\@tabfninsert}
\newlength{\@tabfnwidth}
\newcommand{\tabfootnote}[2]{%
  \setlength{\@tabfninsert}{0.8em}
  \setlength{\@tabfnwidth}{\textwidth}
  \addtolength{\@tabfnwidth}{-\@tabfninsert}
  \addtolength{\@tabfnwidth}{-0.4em}
  \noindent\makebox[\@tabfninsert][r]{\footnotesize$^{#1}$\hfil}\hfill%
  \parbox[t]{\@tabfnwidth}{\footnotesize #2\hfill}}
\catcode`\@=12 

 {\end{list}}
 {\end{list}}
 {\end{list}}

\catcode`\@=11 
\newlength{\@fninsert}
\newlength{\@fnwidth}
\renewcommand{\@makefntext}[1]%
  {\noindent\makebox[\@fninsert][r]{\@makefnmark}\hfil%
  \parbox[t]{\@fnwidth}{#1}}
\catcode`\@=12 
\newlength{\localtextwidth}
\catcode`\@=11 
\newsavebox{\tmpbox}
\newlength{\@captionmargin}
\newlength{\@captionwidth}
\newlength{\@captionitemtextsep}
\renewcommand{\@makecaption}[2]%
  {%
   \vspace{10.pt}
   \setlength{\@captionwidth}{\localtextwidth}
   \addtolength{\@captionwidth}{-\@captionmargin}
   \sbox{\tmpbox}{{\bf #1:}{\it #2}}%
   \ifthenelse{\lengthtest{\wd\tmpbox > \@captionwidth}}%
   {\centerline{\parbox[t]{\@captionwidth}%
   {\tolerance=2000\normalsize%
    {\rm #1:}\hspace{\@captionitemtextsep}{\rm #2}}}}%
   {\centerline{{\bf #1:}\kern1.em{\it #2}}}}
\catcode`\@=12 
\catcode`\@=11 
\renewcommand\section{\@startsection{section}{1}{\z@}%
                                   {-3.5ex \@plus -1ex \@minus -.2ex}%
                                   {2.3ex \@plus.2ex}%
                                   {\normalfont\Large\bfseries}}
\renewcommand\subsection{\@startsection{subsection}{2}{\z@}%
                                   {-3.25ex\@plus -1ex \@minus -.2ex}%
                                   {1.5ex \@plus .2ex}%
                                   {\normalfont\large\bfseries}}
\renewcommand\subsubsection{\@startsection{subsubsection}{3}{\z@}%
                                   {-3.25ex\@plus -1ex \@minus -.2ex}%
                                   {1.5ex \@plus .2ex}%
                                   {\normalfont\large\bfseries}}
\renewcommand\paragraph{\@startsection{paragraph}{4}{\z@}%
                                   {3.25ex \@plus1ex \@minus.2ex}%
                                   {1.2ex \@plus .2ex}%
                                   {\normalfont\normalsize\bfseries}}
\catcode`\@=12 

\newsavebox{\sesbox}
\newlength{\seslen}

\newcommand{\yJB}{\mbox{$y_\mathrm{JB}$}}
\newcommand{\pt}{\mbox{$p_T$}}

\usepackage{graphics}
\usepackage{amsmath}
\usepackage{psfig}  \usepackage{epsfig}

\newcommand{\sleq} {\;\raisebox{-.6ex}{${\textstyle\stackrel{<}{\sim}}$}\;}

\newcommand{\Als}{\mbox{$\alpha_{s}$}}
\newcommand{\Alsmz}{\mbox{$\alpha_{s}(M_Z)$}}
\newcommand{\albar}{\mbox{$\overline{\alpha_{0}}$}}
\newcommand{\albarmuI}{\mbox{$\overline{\alpha_{0}}(\mu_{I})$}}
\newcommand{\error}{uncertainty}
\newcommand{\errors}{uncertainties}
\def\be{\begin{equation}}
\def\ee{\end{equation}}
\def\bea{\begin{eqnarray}}
\def\eea{\end{eqnarray}}
\begin{document}
\selectlanguage{english}


\begin{titlepage}
\thispagestyle{empty}
\title{\vspace*{-4cm}
{\normalsize
\hspace*{\fill}DESY-02-198\\
\hspace*{\fill}November 2002\\[4cm]
}
\bf Measurement of event shapes in deep inelastic 
               scattering at HERA \\}
\author{{\large ZEUS Collaboration}\\[9mm] }
\date{}
\maketitle
\begin{abstract}
      Inclusive event-shape variables have been measured in the
current region of the Breit frame for neutral current deep inelastic
$ep$ scattering using an integrated luminosity of $45.0 {\ {\rm
pb^{-1}}}$ collected with the ZEUS detector at HERA. The variables
studied included thrust, jet broadening and invariant jet mass. The
kinematic range covered was $10 < Q^2 < 20\,480\gev^2$ and $6\times 10^{-4} < x <
0.6$, where $Q^2$ is the virtuality of the exchanged boson and $x$ is
the Bjorken variable.  The $Q$ dependence of the shape variables has
been used in conjunction with NLO perturbative calculations and the
Dokshitzer-Webber non-perturbative corrections (`power corrections')
to investigate the validity of this approach.
\end{abstract}
\end{titlepage}

%
%
%
%

\topmargin-1.cm                                                                                    
\evensidemargin-0.3cm                                                                              
\oddsidemargin-0.3cm                                                                               
\textwidth 16.cm                                                                                   
\textheight 680pt                                                                                  
\parindent0.cm                                                                                     
\parskip0.3cm plus0.05cm minus0.05cm                                                               
\def\3{\ss}                                                                                        
\newcommand{\address}{ }                                                                           
\pagenumbering{Roman}                                                                              
                                                   %
\begin{center}                                                                                     
{                      \Large  The ZEUS Collaboration              }                               
\end{center}                                                                                       
  S.~Chekanov,                                                                                     
  D.~Krakauer,                                                                                     
  J.H.~Loizides$^{   1}$,                                                                          
  S.~Magill,                                                                                       
  B.~Musgrave,                                                                                     
  J.~Repond,                                                                                       
  R.~Yoshida\\                                                                                     
 {\it Argonne National Laboratory, Argonne, Illinois 60439-4815}~$^{n}$                            
\par \filbreak                                                                                     
  M.C.K.~Mattingly \\                                                                              
 {\it Andrews University, Berrien Springs, Michigan 49104-0380}                                    
\par \filbreak                                                                                     
  P.~Antonioli,                                                                                    
  G.~Bari,                                                                                         
  M.~Basile,                                                                                       
  L.~Bellagamba,                                                                                   
  D.~Boscherini,                                                                                   
  A.~Bruni,                                                                                        
  G.~Bruni,                                                                                        
  G.~Cara~Romeo,                                                                                   
  L.~Cifarelli,                                                                                    
  F.~Cindolo,                                                                                      
  A.~Contin,                                                                                       
  M.~Corradi,                                                                                      
  S.~De~Pasquale,                                                                                  
  P.~Giusti,                                                                                       
  G.~Iacobucci,                                                                                    
  A.~Margotti,                                                                                     
  R.~Nania,                                                                                        
  F.~Palmonari,                                                                                    
  A.~Pesci,                                                                                        
  G.~Sartorelli,                                                                                   
  A.~Zichichi  \\                                                                                  
  {\it University and INFN Bologna, Bologna, Italy}~$^{e}$
% ...
\newpage                                                                                           
$^{\    1}$ also affiliated with University College London \\                                      
% ...
                                                           %
                                                           %
\newpage   
                                                           %
                                                           %
\begin{tabular}[h]{rp{14cm}}                                                                       
$^{a}$ &  supported by the Natural Sciences and Engineering Research                               
          Council of Canada (NSERC) \\                                                             
$^{b}$ &  supported by the German Federal Ministry for Education and                               
          Research (BMBF), under contract numbers HZ1GUA 2, HZ1GUB 0, HZ1PDA 5, HZ1VFA 5\\         
$^{c}$ &  supported by the MINERVA Gesellschaft f\"ur Forschung GmbH, the                          
          Israel Science Foundation, the U.S.-Israel Binational Science                            
          Foundation and the Benozyio Center                                                       
          for High Energy Physics\\                                                                
$^{d}$ &  supported by the German-Israeli Foundation and the Israel Science                        
          Foundation\\                                                                             
$^{e}$ &  supported by the Italian National Institute for Nuclear Physics (INFN) \\                
$^{f}$ &  supported by the Japanese Ministry of Education, Science and                             
          Culture (the Monbusho) and its grants for Scientific Research\\                          
$^{g}$ &  supported by the Korean Ministry of Education and Korea Science                          
          and Engineering Foundation\\                                                             
$^{h}$ &  supported by the Netherlands Foundation for Research on Matter (FOM)\\                   
$^{i}$ &  supported by the Polish State Committee for Scientific Research,                         
          grant no. 620/E-77/SPUB-M/DESY/P-03/DZ 247/2000-2002\\                                   
$^{j}$ &  partially supported by the German Federal Ministry for Education                         
          and Research (BMBF)\\                                                                    
$^{k}$ &  supported by the Fund for Fundamental Research of Russian Ministry                       
          for Science and Edu\-cation and by the German Federal Ministry for                       
          Education and Research (BMBF)\\                                                          
$^{l}$ &  supported by the Spanish Ministry of Education and Science                               
          through funds provided by CICYT\\                                                        
$^{m}$ &  supported by the Particle Physics and Astronomy Research Council, UK\\                   
$^{n}$ &  supported by the US Department of Energy\\                                               
$^{o}$ &  supported by the US National Science Foundation\\                                        
$^{p}$ &  supported by the Polish State Committee for Scientific Research,                         
          grant no. 112/E-356/SPUB-M/DESY/P-03/DZ 301/2000-2002, 2 P03B 13922\\                    
$^{q}$ &  supported by the Polish State Committee for Scientific Research,                         
          grant no. 115/E-343/SPUB-M/DESY/P-03/DZ 121/2001-2002, 2 P03B 07022\\                    
\end{tabular}                                                                                      
                                                           %
                                                           %
 
\newpage
\pagenumbering{arabic}                                                         \setcounter{page}{1}
  
\section{Introduction}  \label{sec:intro}
The hadronic final states formed in $e^{+}e^{-}$ annihilation and in
deep inelastic scattering (DIS) can be characterised by a number of
variables that describe the shape of the event.  These variables are
also defined at the parton level, where they are calculable using
perturbative QCD (pQCD).  A comparison of the hadron-level
measurements with the parton-level calculations tests pQCD theory as
well as QCD-based models.  It is critical, however, that the
non-perturbative effects due to hadronisation are correctly taken into
account.

A phenomenological determination of the hadronisation corrections can
be obtained through the use of Monte Carlo (MC) models.  However, an
analytic method has been presented by Dokshitzer and
Webber~\cite{power1} which allows the necessary corrections to be
explicitly evaluated~\cite{power4}.  The mean value of a given shape
variable is taken to be the sum of two parts, one of which is
calculable perturbatively in QCD, while the other models the soft,
non-perturbative contribution.  The measured shape variable then
depends on two experimentally determined constants, namely the QCD
coupling parameter, \Als, and an effective non-perturbative coupling,
\albar.  The non-perturbative contribution is an analytic expression, 
known as a power correction \cite{webber}, which varies inversely with
the hard interaction scale of the event.  In $e^{+}e^{-}$
annihilation, this scale is taken to be $\sqrt{s}$, the centre-of-mass
energy of the incoming particles, while in DIS it is taken to be $Q$,
the square root of the virtuality of the exchanged boson.  The power
corrections provide a potentially powerful tool in the experimental
study of parton physics using perturbative QCD.  They give a good
description of event shapes in $e^{+}e^{-}$
annihilation~\cite{delphi,eventshape1,eeshape}.  A similar success was 
anticipated in DIS, assuming the universality of quark fragmentation,
and hence of event-shape properties.  The first such studies have been
reported by the H1 Collaboration~\cite{H1result}.

 In $e^{+}e^{-}$ annihilation, the event shapes may be evaluated in
the laboratory frame.  In order to study quark fragmentation in DIS, a
frame that isolates the current-quark region of the event from the
proton-remnant region is required, since only the current-quark region
is of interest.  A natural frame for this purpose is the Breit
frame~\cite{breit}.  In this frame, the longitudinal axis is
the direction of the incoming proton, and the current quark
emerges in the opposite direction; the final-state particles are
assigned to the current region if their longitudinal momentum
component is negative, in which case they are interpreted as products
of the hadronisation of the current quark.

This paper presents measurements of event shapes in DIS.  The validity
of the power correction method is studied by examining whether the
data can be correctly described by the theoretical expressions, and by
checking the consistency of the values of \Als\ and \albar\ obtained
from fits to the different event-shape variables.  Measurements are
given of the mean values of the
selected event-shape variables, evaluated in the kinematic range $
6\times 10^{-4} < x < 0.6 $, $ 10 < Q^2 < 20\,480\gev^2$ and $ 0.04 < y
< 0.95 $.  Here $x$ is the Bjorken variable and $y = Q^2/sx$.
Definitions of the event shapes are given in the following section.
The $Q$ dependence of the means of the event-shape variables is
fitted to next-to-leading-order (NLO) estimates from
pQCD~\cite{dis++,disent}, using the Dokshitzer-Webber power
corrections, to determine values of \Als\ and \albar.

\section{Event-shape variables}  \label{sec:esv}

    The event-shape variables studied here  are thrust, jet
broadening, the invariant jet mass and the $C$-parameter. 

   Thrust measures the longitudinal collimation of a given hadronic system,
while broadening measures the complementary aspect. These two
parameters are specified re\-la\-tive to a chosen axis, denoted by a unit
vector $\mathbf{n}$.  Thus:
\begin{equation} T= \frac{\sum_{i} \left|
\mathbf{p}_{i}\cdot\mathbf{n} \right|} {\sum_{i} \left| \mathbf{p}_{i} \right|} =
\frac{\sum_{i} |p_{||}|} {\sum_{i} \left| \mathbf{p}_{i} \right|};
\nonumber \end{equation} \begin{equation} B =
\frac{\sum_{i} \left| \mathbf{p}_{i}\times \mathbf{n} \right|} {\sum_{i} \left|
\mathbf{p}_{i} \right|} = \frac{\sum_{i} p_{\bot}} {\sum_{i} \left| \mathbf{p}_{i}
\right|}.  \nonumber \end{equation} 
The sums in the formulae are
taken over all particles in the chosen region of the event,
namely the current region of the Breit frame. 
With $\mathbf{n}$ taken to be the virtual-photon direction, thrust and
broadening are denoted by $T_\gamma$ and $B_\gamma$, respectively.
Alternatively, both quantities may be measured with respect to the
thrust axis, defined as that along which the thrust is maximised by a
suitable choice of $\mathbf{n}$.  In this case, the thrust and broadening
are denoted by $T_{T}$ and $B_{T}$.

The normalised jet invariant mass, $M$, is defined by 
\begin{equation} M^2 = \frac{\left( \sum_{i} {p_{i}^{\;\mu}}
      \right)^{2}} {\left(2\sum_{i} E_{i}\right)^{2}}.  \nonumber
      \end{equation} 

The $C$-parameter is given by   
\begin{equation} C =
      3\left( \lambda_{1}\lambda_{2} + \lambda_{2}\lambda_{3} +
      \lambda_{3}\lambda_{1} \right), \label{eqn:esv4} \end{equation}
the cyclic sum of the products of the eigenvalues of the
linearized momentum tensor, $\Theta^{\alpha \beta}$, where 
\begin{equation} \Theta^{\alpha \beta} =
      \frac{\sum_{i}\left(\mathbf{p}_{i}^{\;\alpha}
     \mathbf{p}_{i}^{\;\beta} \right)/\left| \mathbf{p}_{i}\right|}{\sum_{i}\left|
      \mathbf{p}_{i}\right| }.  \nonumber
\end{equation} 
Taking $\theta_{ij}$ as the angle between two particles,
Eq.\ (\ref{eqn:esv4}) can be simplified to
\begin{equation}
    C = \frac{3\sum_{ij}|\mathbf{p_{i}}|\,|\mathbf{p_{j}}|\sin^{2}{\theta_{ij}}}
                   {2\left(\sum_{i}\left|\mathbf{p_{i}}\right|\right)^2}.
      \nonumber
\end{equation}

     As seen from the above equations, the shape parameters in the
present study are normalised to the energy in the current hemisphere.
With this normalisation, to ensure infra-red safety, it is necessary
to exclude events in which the energy in the current hemisphere is
less than a certain limit, $\mathcal{E}_{lim}$.  Values of
$\mathcal{E}_{lim}$ of $0.1Q$ and $0.25Q$ have been used.  The primary
analysis is based on event shapes calculated in the $P$-scheme, i.e.\
with particles taken to have zero mass after transformation to the
Breit frame. The $E$-scheme, in which particle masses are
assumed~\cite{scheme}, was used as a cross-check.
  
In the Born approximation, $T_{\gamma}$ and $T_{T}$ are
unity. Consequently, the shape variables $(1 - T_{\gamma})$
and $(1 - T_{T})$ are employed so that 
non-zero values at the parton level are a direct
indicator of higher-order QCD effects.


\section{Detector description and event selection}

The data sample presented here was collected with the ZEUS detector
during 1995-1997 and corresponds to an integrated luminosity of
$45.0\pm 0.7 {\ {\rm pb^{-1}}}$. During this period,
HERA operated with protons of energy $E_p = 820\gev$ and positrons of
energy $E_e = 27.5\gev$.

The ZEUS detector is described in detail elsewhere~\cite{zdet}.  The
main components used in the present analysis are the
uranium-scintillator calorimeter (CAL)~\cite{CAL} and the central
tracking chamber (CTD)~\cite{CTD}, which are positioned in a 1.43~T
solenoidal magnetic field.  The CAL is divided into forward, barrel
and rear sections,\footnote {The ZEUS coordinate system is a
right-handed Cartesian system, with the $Z$ axis pointing in the
proton beam direction, referred to as the `forward direction', and the
$X$ axis pointing left towards the centre of HERA. The coordinate
origin is at the nominal interaction point. The laboratory polar
angle, $\theta$, is measured with respect to the proton beam
direction.}  each of which is subdivided into cells whose energy
deposits are read out independently.  The relative energy resolution,
as measured in test beams, is $18\%/\sqrt{E\mbox{(\gev)}}$ and
$35\%/\sqrt{E(\mbox{\gev})}$ for electrons and hadrons,
respectively. The interaction vertex was measured using the CTD with a
typical resolution of 0.4 (0.1) cm along (transverse to) the beam
direction.

The DIS kinematic variables and
the four-vector of the virtual photon were reconstructed using the
double-angle (DA) method~\cite{dbang}.
The following additional experimental quantities are defined:
\begin{itemize}
\item $\delta = \Sigma_i E_i (1-\cos\theta_i)$, where $E_i$ is the
energy measured by a calorimeter cell, and $ \theta_i$ is the polar
angle of the cell relative to the vertex position.  The sum
runs over all calorimeter cells;
\item $\yJB = \Sigma_i E_i (1-\cos\theta_i)/2E_e$, where the sum
runs over all calorimeter cells except those associated with the
scattered positron;
\item $y_e = 1 - ({E'_e}/{2E_e})(1-\cos\theta_e)$, where $E'_e$ and
$\theta_e$ are the energy and angle of the scattered positron.
\end{itemize}

A pure sample of DIS events in a well-defined kinematic region was selected as follows:
\begin{itemize}
\item the events must pass the trigger, whose critical
component in this case was a selection on a high-energy scattered
positron identified in the CAL;
\item a primary vertex must be reconstructed, with one or more
tracks, satisfying $-50 < Z_{\mathrm{vtx}} < 40$ cm;
\item a well-identified scattered positron, which was found by means
of a neural-network procedure~\cite{sinistra},  
must satisfy $E'_{e} \geq10\gev$; 
\item the measured impact position  of the scattered positron on the
face of the rear calorimeter must be outside a square of $ 16 \times
16 \mathrm{cm}^2 $ centred on the beampipe; 
\item $35< \delta <60\gev$, to remove photoproduction events
where the scattered positron is lost down the beampipe, and also
to reduce the effects of initial-state radiation;
\item $\yJB \geq 0.04$, to ensure a well-measured hadronic system;
\item $ y_e\,\leq\,0.95$, to reduce  background
contamination from photoproduction~\cite{nophoto};
\item $Q^{2} \geq 10\gev^2$ and $x > 6\times 10^{-4} $.  \end{itemize}

Both track and calorimeter information were used to determine the
event shapes.  Calorimeter cells were first grouped to form clusters
and these clusters were then associated with tracks, where possible,
to form energy-flow objects, EFO's, associated with the hadrons formed
in the interaction~\cite{zufos}.  Both tracks and clusters were
required to have $20^\circ < \theta < 160^\circ$ and 
transverse momentum $\pt > 150\mev$.  At this stage, EFO's with tracks
were assigned the mass of the pion, while those without tracks,
corresponding mainly to photons from $\pi^0$ decays, were assigned
zero mass.  Each accepted EFO was then transformed to the Breit frame,
where it was assigned to the current region if its longitudinal momentum 
was negative.  Subsequently, the masses were assigned
according to the $P$- or $E$-scheme.  A total of
321\,000 events resulted from these selections and  were used in the analysis.

\section{QCD models and event simulation}  \nonumber 
Monte Carlo  event simulation was used to correct
the data for acceptance and resolution effects.  The detector
simulation was performed with the GEANT~3.13 program~\cite{GEANT}.

Neutral current DIS events were generated using the DJANGOH~1.1
package~\cite{django}, combining the LEPTO 6.5.1~\cite{LEPTO}
generator with the HERACLES~4.6.1 program~\cite{HERACLES}, which
incorporates first-order electroweak corrections.  The parton cascade
was modelled with the colour-dipole model (CDM), using the
ARIADNE~4.08 \cite{ariadne,highqari} program. In this model, coherence
effects are implicitly included in the formalism of the parton
cascade.  The Lund string-fragmentation model \cite{string} is used
for the hadronisation phase, as implemented in JETSET~7.4
\cite{pythiajetset}.

Further samples were generated with the HERWIG~5.9 program
\cite{herwig}, which does not apply electroweak radiative corrections.
The coherence effects in the final-state cascade are included by
angular ordering of successive parton emissions, and a clustering model
is used for the hadronisation \cite{cluster}. Events were also
generated using the MEPS option of LEPTO within DJANGOH, which
subsequently uses a parton-showering model similar to HERWIG.  To
achieve agreement with the data, a diffractive component of $14\%$ of
the DIS events~\cite{thesis} was required and was simulated using
RAPGAP~2.08~\cite{rapgap}.

For ARIADNE, the default parameters were used.  The LEPTO simulation
was run with soft-colour interactions turned off, and HERWIG was
retuned\footnote{The parameter PSPLT was set equal to 1.8; otherwise
default parameters were used.} to give closer agreement with the
measured shape variables at low $Q$; the CTEQ4D~\cite{cteq4m}
parameterisations of the proton parton distribution functions (PDF)
were taken.  The Monte Carlo event samples were passed through
reconstruction and selection procedures identical to those for the
data.

\section{Data correction}

The event shapes were evaluated for event samples in selected bins of
$x$ and $Q^2$.  The choice of the bin sizes\cite{breit1} was motivated
by the need to have good statistics while keeping  the
migrations, both between bins, and from the current to the target region
within each bin, to a minimum.  The kinematic bin boundaries are listed in
Table~\ref{table:kinematics}.

In each ($x$, $Q^2$) bin, the ARIADNE MC was used to investigate the
event acceptance and the acceptance in each bin of the event-shape
variable.  The acceptance was defined at the hadron level as the
ratio of the number of reconstructed and selected events to the number
of generated events in the given bin.  The event acceptance exceeded
$70\%$ for all bins, while the event-shape acceptance was less than
$70\%$ only at the extremes of the $Q^2$ range and at low $y$.

Agreement was found between the uncorrected data and the predictions
of ARIADNE throughout the entire kinematic range of each event-shape
variable, thus confirming its suitability for the purposes of
correcting the data.  Good agreement with ARIADNE was also found for
the energy-flow~\cite{zeus:efl} and charged-track
distributions~\cite{janethesis} studied in previous analyses.  The
data were also compared with the HERWIG predictions; here the
agreement with data was satisfactory but less good than for ARIADNE.

  The correction factors for the means of the shape variables were
evaluated as the ratios of the generated to the observed values of the
mean in each ($x,Q^2$) bin.  These correction factors, which
were used for the subsequent analysis, lie within the range 0.75 -
1.12 and are typically within $\pm5\%$ of unity.  As a check, the
calculation was repeated correcting the individual bins using the
acceptances as described above.

  The generated distributions include the products of strong and
electromagnetic decays, together with $K^0_S$ and $\Lambda$ decays,
but exclude the decay products of weakly decaying particles with
lifetime greater than \mbox{$3\times10^{-10}$s}.  The correction
procedure accounts for event migration between ($x,Q^2$) intervals,
QED radiative effects, EFO-reconstruction efficiency and energy
resolution, acceptances in \pt\ and $\theta$, EFO migration between
the current and target regions, and the decay products of $K^{0}_S$
and $\Lambda$ decays that were assigned to the primary vertex.

\section{Systematic checks} \label{sec:sys} 

The systematic \errors\ in the measurement can be divided into three
types, due to the MC model used, to the event
reconstruction and selection, and to the EFO reconstruction. The
systematic checks were as follows:
\begin{itemize} 
\item the data were corrected using a different hadronisation and
parton-shower model, namely HERWIG or LEPTO, in place of ARIADNE;
\item the cut on \yJB\ was increased to 0.05;
\item the cut on $\delta$ was changed to $40 < \delta< 60$ GeV; this
harder cut estimates any residual uncertainties in the
photoproduction background;
\item the double-angle kinematics were recalculated after removing
CAL deposits due to backwardly scattered particles (albedo) from the material close to
the proton beampipe \cite{thesis,nc};
\item the measured energies of clusters in the calorimeter were varied
by $\pm3\%$, $\pm1\%$ and $\pm2\%$ for the forward, barrel and rear
CAL sections, respectively, corresponding to the uncertainties on the
associated energy scales;
\item the EFO cuts at $\theta = [20^\circ, 160^\circ] $ and $\pt > 150
\mev$  were tightened to $\theta = [25^\circ, 155^\circ] $ and $\pt >
200\mev$; the cuts were also removed;
\end{itemize} 
The largest systematic \error\ was due to using HERWIG as the
hadronisation model.  The LEPTO model produced smaller changes.
The other systematics were typically at the level of $\sleq1\%$, smaller
than or similar to the statistical errors.

\section{Power corrections} \label{sec:power} 
Next-to-leading-order QCD calculations of the shape parameters have
been made using the programs DISASTER++~\cite{dis++} and
DISENT\cite{disent}, which give parton-level distributions.  Both
programs used the CTEQ4A PDFs.  To determine the theoretical
$\alpha_{s}$ dependence of the variables, high-statistics event
samples were generated for each of five values of
$\alpha_{s}(M_Z)$~\cite{cteq4m}.  For both NLO calculations, the mean
value of each shape variable was found to be linearly dependent on
\Als\ in the appropriate range.  For each bin, therefore, the
calculated value of the shape variable may be used to estimate the
value of \Als\ at the chosen reference scale, namely the mass of the
$Z^0$ boson.

The relationship used to calculate $\alpha_{s}(Q)$ in terms of
\Alsmz\ is~\cite{runninga}
\begin{equation} 
\alpha_{s}(Q) = \frac{\alpha_{s}(M_{z})}{1 +
 \alpha_{s}(M_{z})L^{(n)}\left(\frac{Q}{M_{z}}\right)}, 
 \nonumber \end{equation} where, for the 2-loop form, 
 \begin{equation} L^{(2)}\left(\frac{Q}{M_{z}}\right) =
 \left(\frac{\beta_{0}}{2\pi} +
 \frac{\beta_{1}}{8{\pi^{2}}}\right)\log\left(\frac{Q}{M_{z}}\right),
 \nonumber
\end{equation} 
with 
\begin{equation}
 \beta_{0}=11-\frac{2}{3}N_{f} \;\;\;\mathrm{~and~}\; \beta_{1} = \frac{306
- 38N_{f}}{3}.  \nonumber
\end{equation} 
Here, $N_{f}$ is the number of active quark flavours at the scale 
$M_Z$, taken to be five.

Before comparison with experimental data, the calculated values of the
shape parameters require correction for the effects of hadronisation.
Dokshitzer and Webber calculated power corrections to the 
event-shape variables in $e^{+}e^{-}$ annihilation, assuming
an infrared-regular behaviour of the effective coupling,
$\alpha_\mathrm{eff}$ \cite{power1,power4}.  The
technique was subsequently applied to the case of
DIS~\cite{webber} and has been used here.

In this approach, a constant, \albar,  is introduced,
which is taken to be independent of the shape variable.  This constant is
defined as the first moment of the effective strong coupling below the
scale $\mu_{I}$ and is given by: 
\begin{equation}
\albarmuI = \frac{1}{\mu_{I}} \int_{0}^{\mu_{I}}
      \alpha_{\mathrm{eff}}(\mu)d\mu, \nonumber
\end{equation} 
where the variable $\mu_{I}$ is the lower limit for the perturbative
approach to be valid. This is taken to be $2\gev$, in common with
previous analyses~\cite{delphi,eventshape1,eeshape,H1result}.

The theoretical prediction for a mean event-shape variable,
denoted by $\left< V\right>$, is then given by  
\begin{equation} \left< V\right> =
\left< V\right>_\mathrm{NLO} + \left< V\right>_\mathrm{pow},
\label{eqn:nlo+power}
\end{equation} 
where $\left< V\right>_\mathrm{NLO}$ is calculated perturbatively, 
and $\left< V\right>_\mathrm{pow}$ is the
power correction.  The generalized power correction is given by 
\begin{equation}
\left< V\right>_\mathrm{pow} = a_{V} \frac{4{\cal{M}}A_{1}}{{\pi}Q}.
\label{eqn:power} \end{equation} This has a $1/Q$ dependence with a
calculable coefficient, $a_{V}$.  The variable $\cal{M}$ is the `Milan
 factor' of value 1.49~\cite{milan}, which takes into account two-loop
 corrections; it has a relative uncertainty of about $\pm20\%$, due to
 three- and higher-loop effects. The term $A_{1}$ is given by:
\begin{equation}
     A_{1} = \frac{C_{F}}{\pi}\mu_{I} \left[ \overline{\alpha_{0}} -
            \alpha_{s}(\mu_R) - \frac{\beta_{0}}{2\pi} \left(
            \log{\left(\frac{\mu_R}{\mu_{I}}\right)} +
            \frac{K}{\beta_{0}} + 1 \right) \alpha_{s}^{2}(\mu_R)
            \right] \label{eqn:A_1} \end{equation} 
where $C_{F}=\frac{4}{3}$, $K=\frac{67}{6} - \frac{\pi^{2}}{2} -
\frac{5}{9}N_{f}$ and $N_{f}$ is taken to be five. The central
analysis was performed with the renormalisation scale ${\mu_R}$ set to
$Q$.  The dependence of the perturbative predictions on this scale was
studied by introducing a variable parameter, $x_{\mu}$, such that
${\mu_R} = x_{\mu}Q$.  The NLO calculations, but not the power
corrections, depend also on the factorisation scale, which is varied
from its central value $Q$ by introducing a similar parameter,
$x_{F}$.

  In Eq.~(\ref{eqn:A_1}), the low-energy contribution to the mean shape
variable is determined by \albar, while the remaining terms subtract out
the integral, up to $\mu_{I}$, of the perturbative expression for
the shape average.  Above this limit, the perturbative expression is
taken to be applicable.

The values of $a_V$ for $(1 - T_T),$  $(1 - T_\gamma),$ $C$ 
and $M^2$ are respectively 2, 2, 1 and $3\pi$.  For $B_\gamma$,
the form~\cite{powerjetb} 
\begin{equation} a_V =
\frac{\pi}{2\sqrt{2C_{F}\alpha_\mathrm{CMW}(\overline{Q})}} + 0.75 -
\frac{\beta_{0}}{12C_{F}} + \eta_{0}, \label{eqn:B} 
\end{equation} 
has been used, where $\eta_{0}=-0.614$,
$\overline{Q} = \mu_Re^{-\frac{3}{4}}$, and the physical coupling
$\alpha_\mathrm{CMW}$ is related to the standard $\alpha_{\overline{MS}}$ by
\begin{equation} \alpha_\mathrm{CMW} = \alpha_{\overline{MS}}\left(1 +
K\frac{\alpha_{\overline{MS}}}{2\pi}\right) \nonumber.
\end{equation}
In Eq.\ (\ref{eqn:B}), $\beta_{0}$ was calculated with $N_{f}=3$,
which is appropriate for fragmentation.  In the corresponding
expression for $B_T$, the factors $2C_{F}$ and $12C_{F}$ were replaced
by $C_{F}$ and $6C_{F}$, respectively.  

It has recently been found~\cite{ds2002} that for $B_\gamma$, there is
an additional $x$-dependent term in Eq.\ (\ref{eqn:B}).  The
importance of this term in the present analysis has been found to be
small; for further comments, see below.

\section{Results}
In this section, the observed $Q$ dependences of the mean values of
the event-shape parameters are presented and compared with the
expectations from theory.  To study the sensitivity to
$x$, results for $Q^2 < 320\gev^2$ are calculated separately in the high and low $x$ ranges as well
as for the full $x$ range.  The values of $\mathcal{E}_{lim}$ used were
$0.25Q$, as recommended~\cite{resum} to ensure
convergence of the perturbative series involving
${\ln(Q/{2\mathcal{E}_{lim}})}$, and $0.1Q$~\cite{H1result}.
The mean values of the event-shape parameters for $\mathcal{E}_{lim} = 0.25Q$
are listed in Table~\ref{table:tz.means} together with their statistical and 
systematic \errors.  

\begin{figure} \begin{center}
      \scalebox{1.0}{\epsfig{file=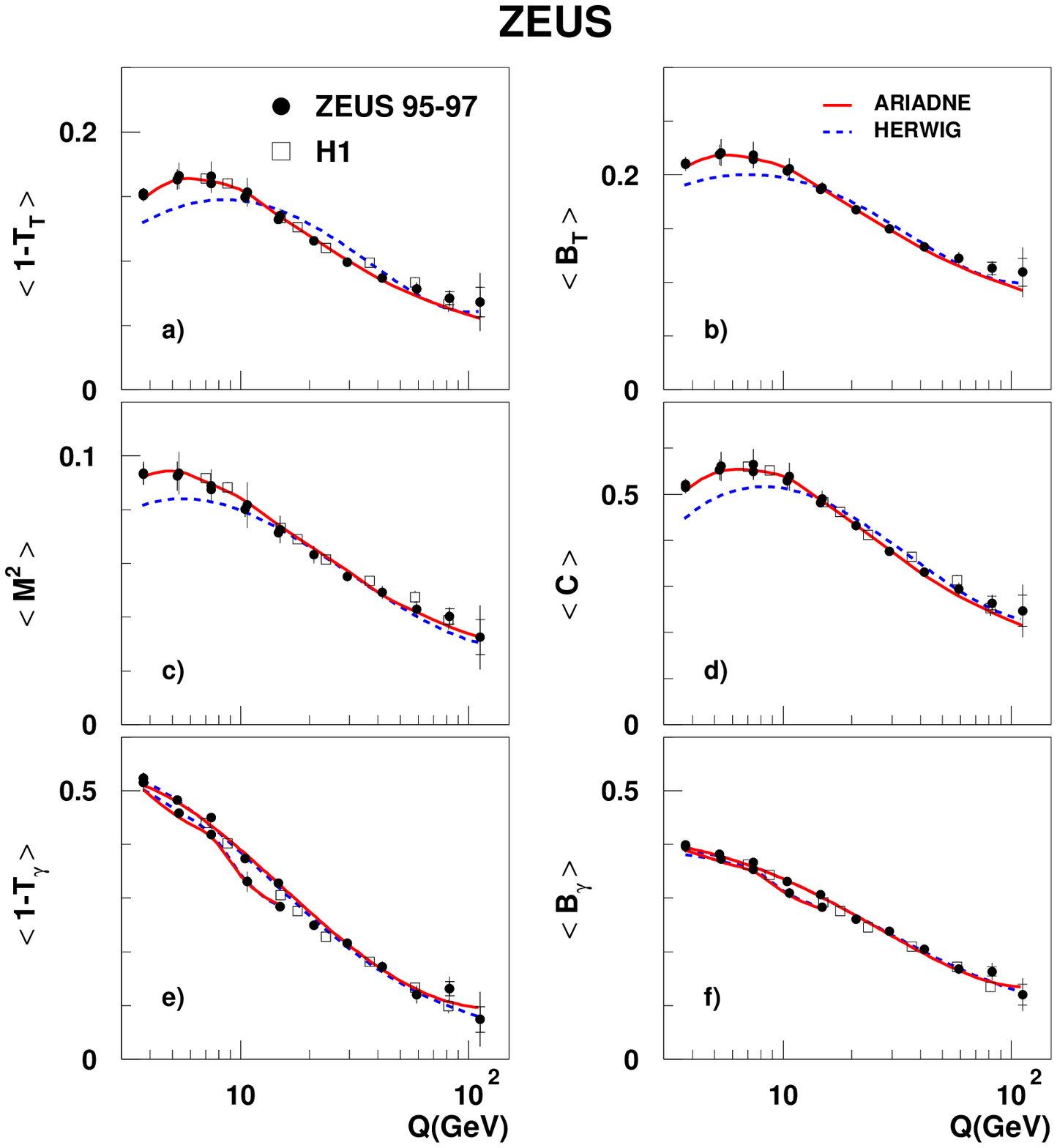,width=15cm,%
      bbllx=0pt,bblly=0pt,bburx=520pt,bbury=600pt,clip=yes}}
      \caption{Comparison of the mean event-shape parameters (solid
      points) with ARIADNE and HERWIG predictions.  The plots refer to
      a) thrust with respect to the thrust axis, b) jet broadening
      with respect to the thrust axis, c) invariant jet-mass squared,
      d) $C$-parameter, e) thrust with respect to the virtual photon
      axis, f) broadening with respect to the virtual photon axis.
      The data were corrected using ARIADNE.  The inner error bars are
      statistical; the outer are statistical plus
      systematic added in quadrature.  The open squares are H1
      data, with statistical uncertainties 
      which are in most cases covered by the
      symbol.  In e) and f), where the $x$ variation is biggest, smoothed
      curves are drawn through the MC points for all $x$ $(Q^2 > 320
      \gev^2)$ and low $x$ $(Q^2 < 320 \gev^2)$, and separately
      through the high-$x$ MC points $(Q^2 < 320 \gev^2)$. A value of
      $\mathcal{E}_{lim}=0.1 Q$ was used.}  \label{fig:1} \end{center}
      \end{figure}

  Figure \ref{fig:1} shows the corrected mean values of the event
shapes as a function of $Q$, using $\mathcal{E}_{lim} = 0.1Q$,
together with the H1 measurements \cite{H1result}, with which there is
good agreement.  The mean values fall with $Q$ at the higher $Q$
values, and at lower $Q$ show an $x$ dependence at fixed $Q$ which is
most pronounced for the variables measured with respect to the photon
axis.  Both here and with $\mathcal{E}_{lim} = 0.25Q$ (not shown),
good agreement is found with the predictions of ARIADNE over the
entire $Q$ range.  The agreement with HERWIG is less good,
particularly for the variables $(1-T_T)$, $B_T$, $M^2$ and $C$ at low
$Q$. Although not used in the subsequent fits, the data for $Q^2 <
80\gev^2$ are included for purposes of completeness, and to allow
comparison with more extended theoretical fits if they become
available.

  The data were fitted to the sum of an NLO term, obtained from
DISASTER++ or DISENT, plus a power correction as described above.  The
theoretical calculation neglects terms of the order of
$1/{Q^{2}}$~\cite{power1}.  Consequently, the analyses have been
confined to the region $Q> 9\gev$, i.e.\ bins 7 - 16 in
Table~\ref{table:kinematics}.  With $\mu_{I}$ fixed at $2\gev$, there
are two parameters in Eq.~(\ref{eqn:nlo+power}), $\alpha_{s}(M_{Z})$
and \albar, that can be varied to obtain the best agreement between
calculation and data.

\begin{figure} \begin{center}
      \epsfig{file=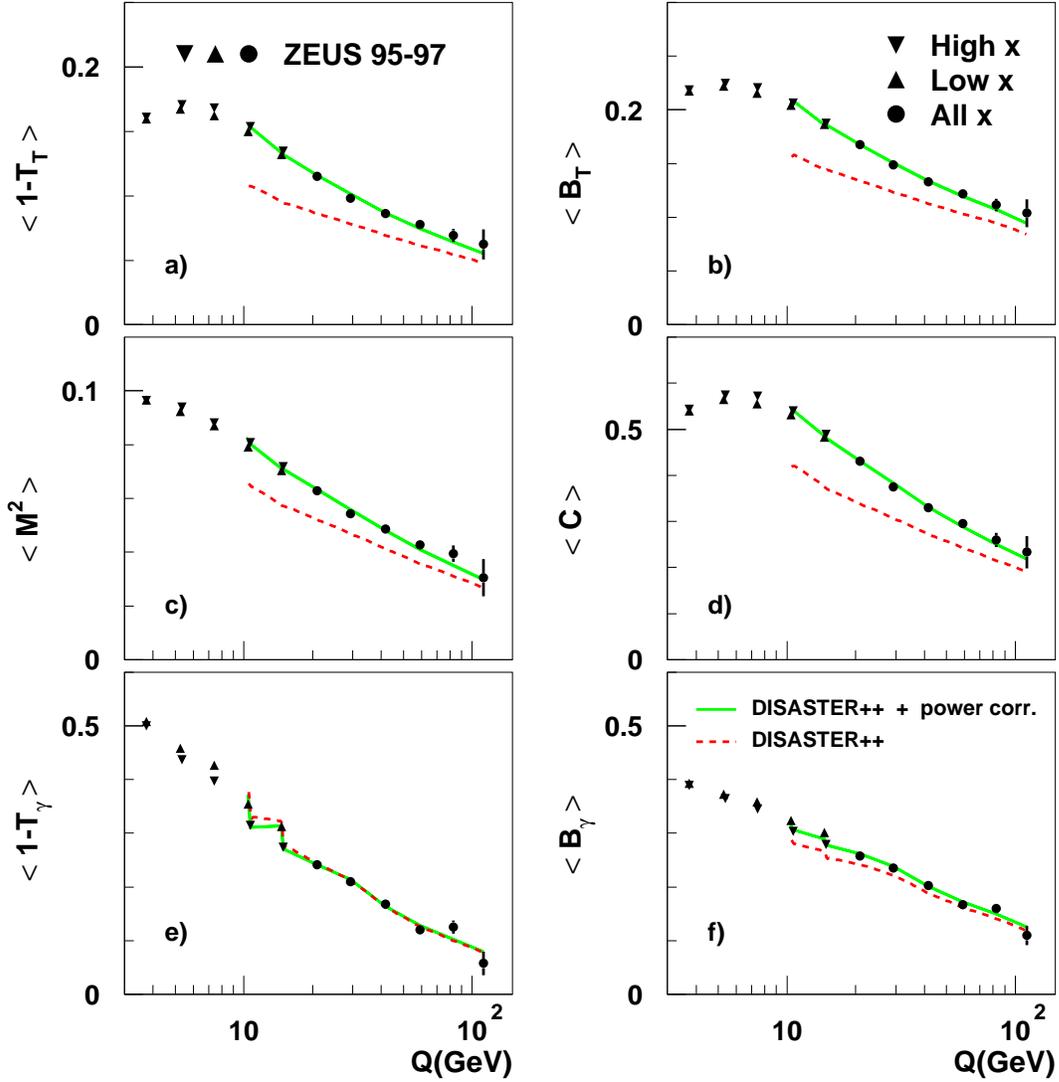,width=16.5cm} \caption{Fits to mean
      values of the shape variables versus $Q$, with
      $\mathcal{E}_{lim} = 0.25Q$.  Plots a) - f) are defined in the
      caption to Fig.\
      \ref{fig:1}.  The lines join the fit values at the $Q$ values of
      the data: the solid line is the fitted NLO prediction from
      DISASTER++ plus the power correction, while the dashed line is
      the fitted DISASTER++ contribution alone. `High $x$' and `low
      $x$' refer to the subdivisions as defined in Table 1; `all $x$'
      refers to the points with $Q > 20 \gev,$ which are not
      subdivided in $x$. }  \label{fig:2} \end{center} \end{figure}

  Two types of fit were studied: the offset method and the Hessian
method. The offset method \cite{offset,zqcd02} uses a $\chi^2$ defined
using a diagonal error matrix with errors given by the statistical
errors on the data points combined in quadrature with the uncorrelated
errors arising from the limited statistics of the DISENT and
DISASTER++ calculations. The Hessian method \cite{hessian,zqcd02} uses
a full error matrix which includes correlated off-diagonal terms due
to the systematic \errors.  The fits obtained using the offset method
are shown in Fig.~\ref{fig:2}.  For the Hessian method, the four
major systematics, namely those associated with $\delta$, the
tightening and relaxation of the EFO angular and \pt\ cuts, and the
use of HERWIG, were included as off-diagonal terms in the error
matrix.  As expected, the $\chi^2$ is reduced with the Hessian method,
and the fitted error, which includes the systematic contribution, is
approximately a factor of two smaller than the systematic uncertainties
estimated from the offset fits.  For the variables $(1-T_T)$, $B_T$,
$M^2$ and $C$, the values of \Als\ from the two fit methods agree
within the statistical uncertainties.  For the \Als\ values from
$(1-T_\gamma)$ and $B_\gamma$, and for all evaluations of \albar, the
two fit methods give results that in general differ by more than the
statistical errors; they do however agree within the offset systematic
\error.  The difference in the results of the fits originates  
primarily from the use of HERWIG in place of ARIADNE.  The Hessian
method relies upon Bayesian priors, specifically the assumption of
Gaussian distributions, for the systematic \errors. There is no reason
to believe that this assumption is correct for the dominating
fragmentation systematic. Consequently, the results presented here are
based on the offset method with its conservative estimate of
systematic effects.

\begin{figure} \begin{center}
      \scalebox{0.84}{\includegraphics{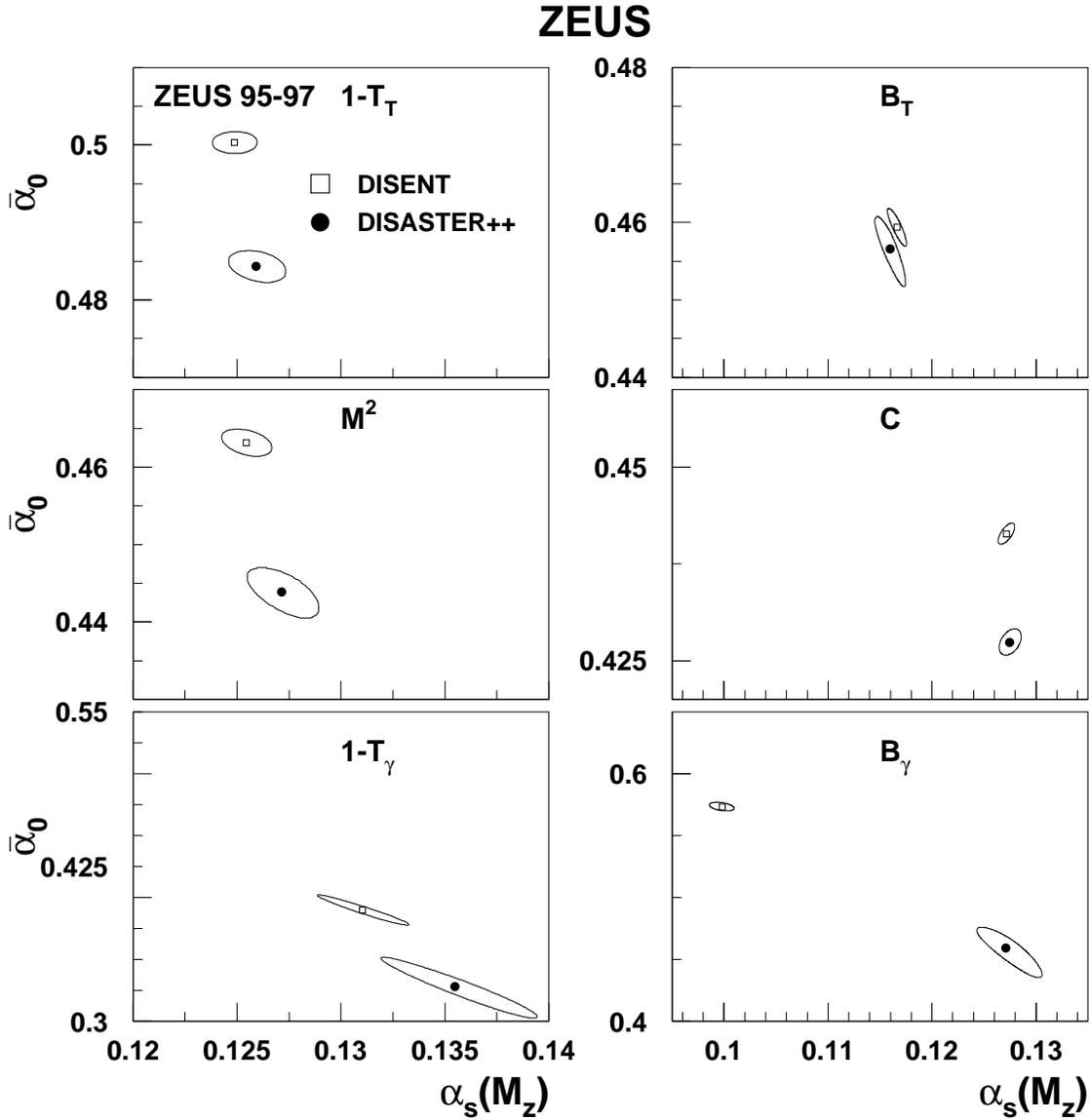}} 
      \caption{Contour plots
      for the parameters \Alsmz\ and \albar\ fitted to the mean values
      of thrust and broadening measured with respect to the photon
      axis, jet-mass squared, $C$-parameter, and thrust and broadening
      measured with respect to the thrust axis.  Results are shown for 
      fits, using all data points for $Q^2> 80\gev^2$, based on
      DISASTER++ and DISENT with $\mathcal{E}_{lim}=0.25 Q$.  The
      contours show the one-standard-deviation limits determined using
      statistical uncertainties only.}
\label{fig:3} \end{center} \end{figure}

The DISASTER++ calculation gives predictions in closer agreement with
analytic calculations~\cite{resum,gavin} than does DISENT, which is
believed to contain errors~\cite{dslatest}.  The fits using DISASTER++
for the NLO term are therefore taken to give the more reliable
estimates of \Als\ and \albar.  The DISENT-based analysis provides a
check and facilitates comparison with H1, whose analysis used this
program.  Reasonable fits are obtained; those based on DISASTER++,
using $\mathcal{E}_{lim} = 0.25Q$, are shown in Fig.\ \ref{fig:2},
while those for DISENT (not shown) are similar.  The fitted
power-correction term is substantial except for the variables that are
based on the virtual-photon axis.  The results of the fits using
DISASTER++ and DISENT are compared in Fig.~\ref{fig:3}, where it can
be seen that, with the exception of $B_{\gamma}$, \Als\ is the same
within statistical errors for the two NLO calculations.  For all
variables, \albar\ determined using DISASTER++ is smaller
than when using DISENT.

As seen in Fig.~\ref{fig:2}, the data have a significant
$x$-dependence at a given $Q$; consequently, fits have been made using
the high-$x$ and low-$x$ selections, as well as to the full set of
used bins.  Results from the DISASTER++ fits are shown in the
contour plots of Fig.~\ref{fig:4}.

\begin{figure} \begin{center}
    \scalebox{0.8}{\includegraphics{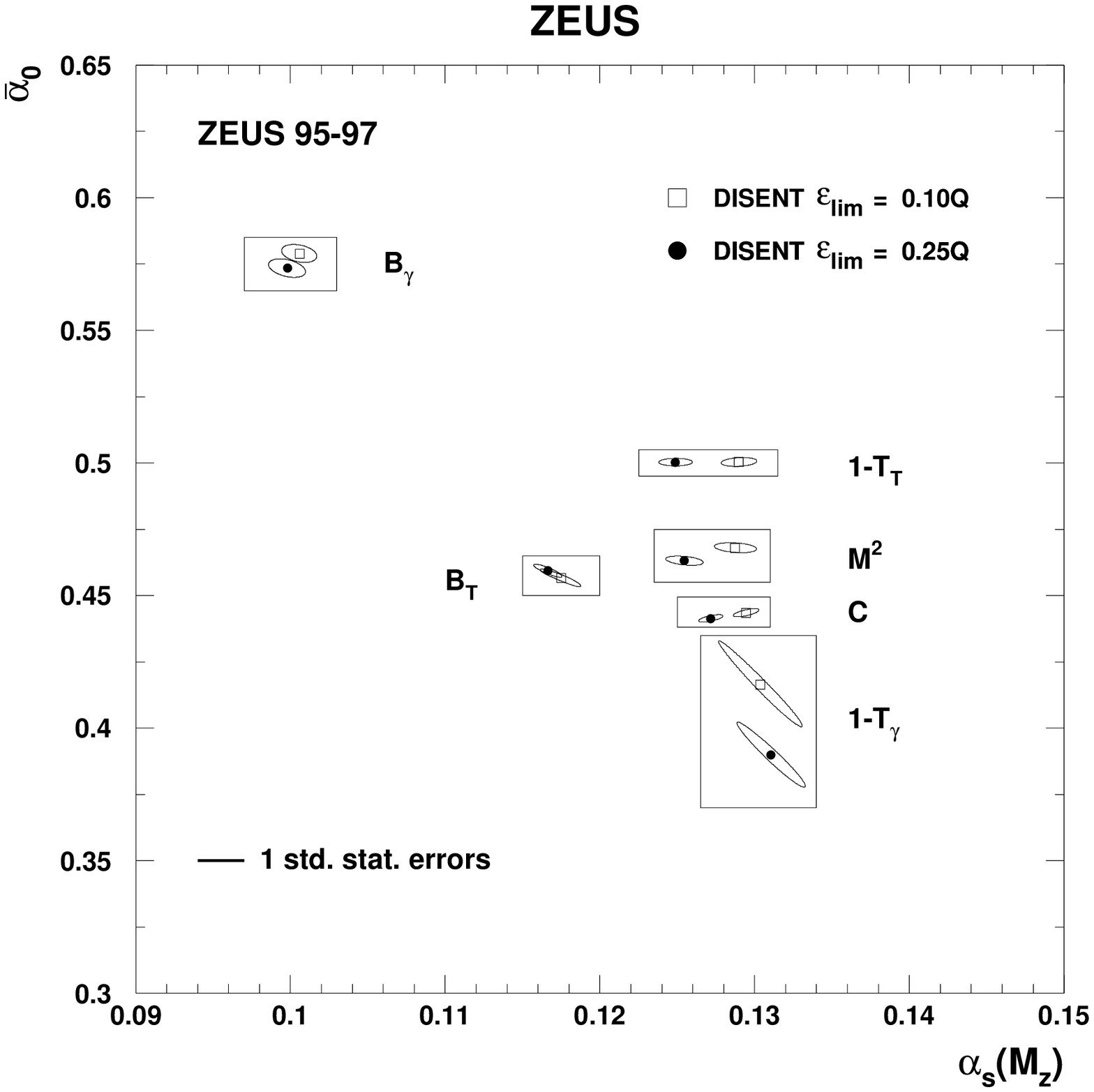}} \caption{Contour plots
    for \Alsmz\ and \albar\ fitted to the mean values of the
    event-shape variables. The fits are based on DISENT, using all
    data for $Q^2> 80\gev^2$, with energy cuts $\mathcal{E}_{lim}=0.1
    Q$ (open squares) and $0.25 Q$ (filled circles).  
    The contours show the one-standard-deviation
    limits determined using statistical uncertainties only. The rectangles
    enclose the associated pairs of points to guide the eye. }
    \label{fig:5} \end{center} \end{figure}

To allow a direct comparison with the H1 results~\cite{H1result},
Fig. \ref{fig:5} shows the determinations of \Als\ and \albar\ using
DISENT and $\mathcal{E}_{lim} = 0.1Q$; as expected from the agreement
of the measured data points~(Fig.~\ref{fig:1}), the agreement in the
fitted \Als\ and \albar\ values is, in general, good.  In particular,
a low value of \Als\ for $B_\gamma$ is confirmed.  The influence of
the $\mathcal{E}_{lim}$ selection is also illustrated in
Fig.~\ref{fig:5}, which shows, for DISENT, that the different
selections lead to \albar\ values that agree within about two standard
deviations.  In contrast, \Als\ appears more sensitive to
$\mathcal{E}_{lim}$ for several of the variables.  In general, it is
found that $\mathcal{E}_{lim} = 0.25Q$ gives a smaller variation of
the fitted \Als\ and \albar\ with $x$ than is found for
$\mathcal{E}_{lim} = 0.1Q$ (not shown).  Given this, together with the
theoretical preference for the higher
$\mathcal{E}_{lim}$~\cite{resum}, the central analysis is based on the
data evaluated with $\mathcal{E}_{lim} = 0.25Q$.

The experimental systematic \errors\ on \Als\ and \albar\ were
estimated by repeating the fits with the systematic variations
described in Section~\ref{sec:sys}.  The largest effect resulted from
correcting the data with HERWIG instead of ARIADNE.  The use of HERWIG
gave a systematic increase in \Als\ and a systematic decrease in
\albar; these shifts are possibly attributable, respectively, to the
use of parton showers rather than the colour-dipole model, and to the
different fragmentation schemes used in the models. Also, HERWIG does
not contain electroweak terms. The variables $(1-T_{\gamma})$ and
$B_{\gamma}$ are in addition sensitive to the method of reconstructing
the kinematic variables, owing to their dependence on the photon
direction.

\begin{figure} \begin{center}
    \scalebox{0.8}{\includegraphics{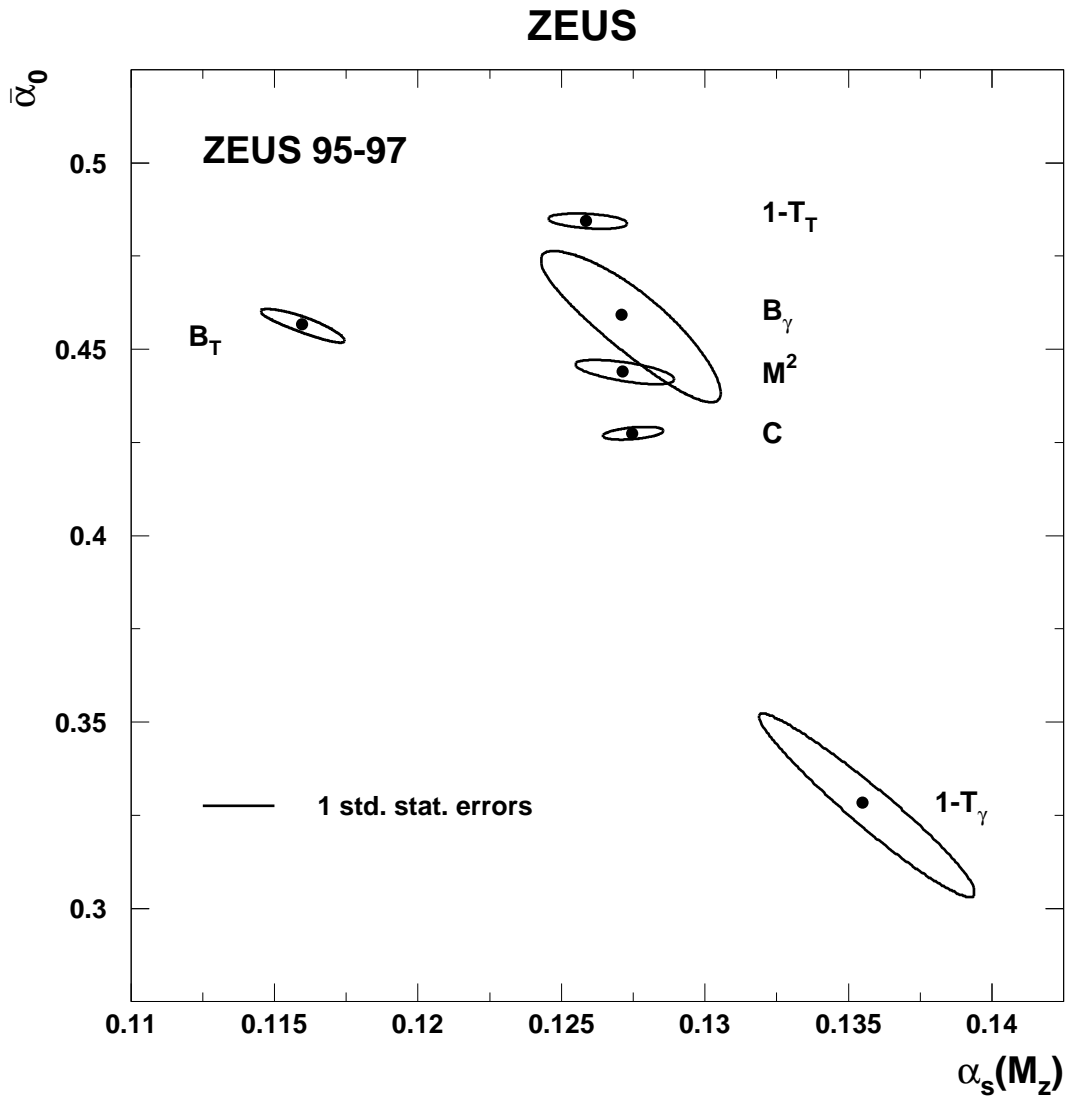}}
    \caption{Contour plots for \Alsmz\ and \albar\ fitted to the mean
    values of the event-shape variables.  The fits are based on
    DISASTER++, with $\mathcal{E}_{lim}=0.25 Q$.  The contours show
    the one-standard-deviation limits determined using statistical
    \errors\ only. For further comments, see text; the full
    systematic errors, which are strongly correlated between the
    different variables, are given in Tables 3 and 4.} \label{fig:6}

\end{center} \end{figure}

Figure ~\ref{fig:6} summarises the \Als\ and \albar\ values from fits
using the statistical errors, in order to indicate the degree of
agreement between the different measurements.  The systematic uncertainties
on the data introduce highly correlated effects on the results
obtained from the different shape variables, and so are not included
here.  The inconsistency which is evident between the different
determinations is discussed below.

To estimate the theoretical uncertainties, the fragmentation and
renormalisation scales were varied by a factor of two, and studies were
made of the effects of changes to $\mu_{I}$ and to the Milan factor.
To give an indication of the uncertainties due to mass effects, the
data were reanalysed using the $E$-scheme~\cite{scheme}.  It was found
that \Als\ depends strongly on $\mu_{I}$, decreasing as $\mu_{I}$ is
increased. If the model were robust, \Als\ should have little
dependence on $\mu_{I}$.

The fit results including experimental and theoretical systematic
\errors\ are collected in Tables \ref{table:finalfitnew1}
and~\ref{table:finalfitnew2}.  The dominant uncertainty comes from the
variation of the renormalis\-a\-tion scale.  The
renormalis\-a\-tion-scale \error\ quoted here follows the procedure
employed in the $e^+e^-$ studies \cite{delphi,eventshape1,eeshape}
using Eq.~(\ref{eqn:power}). If, following H1~\cite{H1result}, $Q$ in
Eq.~(\ref{eqn:power}) were replaced by $\mu_{R}$, the theoretical
\error\ due to the $x_{\mu}$ variation would be approximately a factor
of two larger.  The influence of the $x$-dependent term in Eq.\
(\ref{eqn:B}) was examined using the CTEQ5M proton
structure~\cite{cteq5m}.  While an improved fit to the data was
obtained, the changes to the resulting \Als\ and
\albar\ values were less than their statistical uncertainties. 

\section{Discussion}\label{sec:discussion}

\begin{figure} \begin{center}
      \scalebox{0.84}{\includegraphics{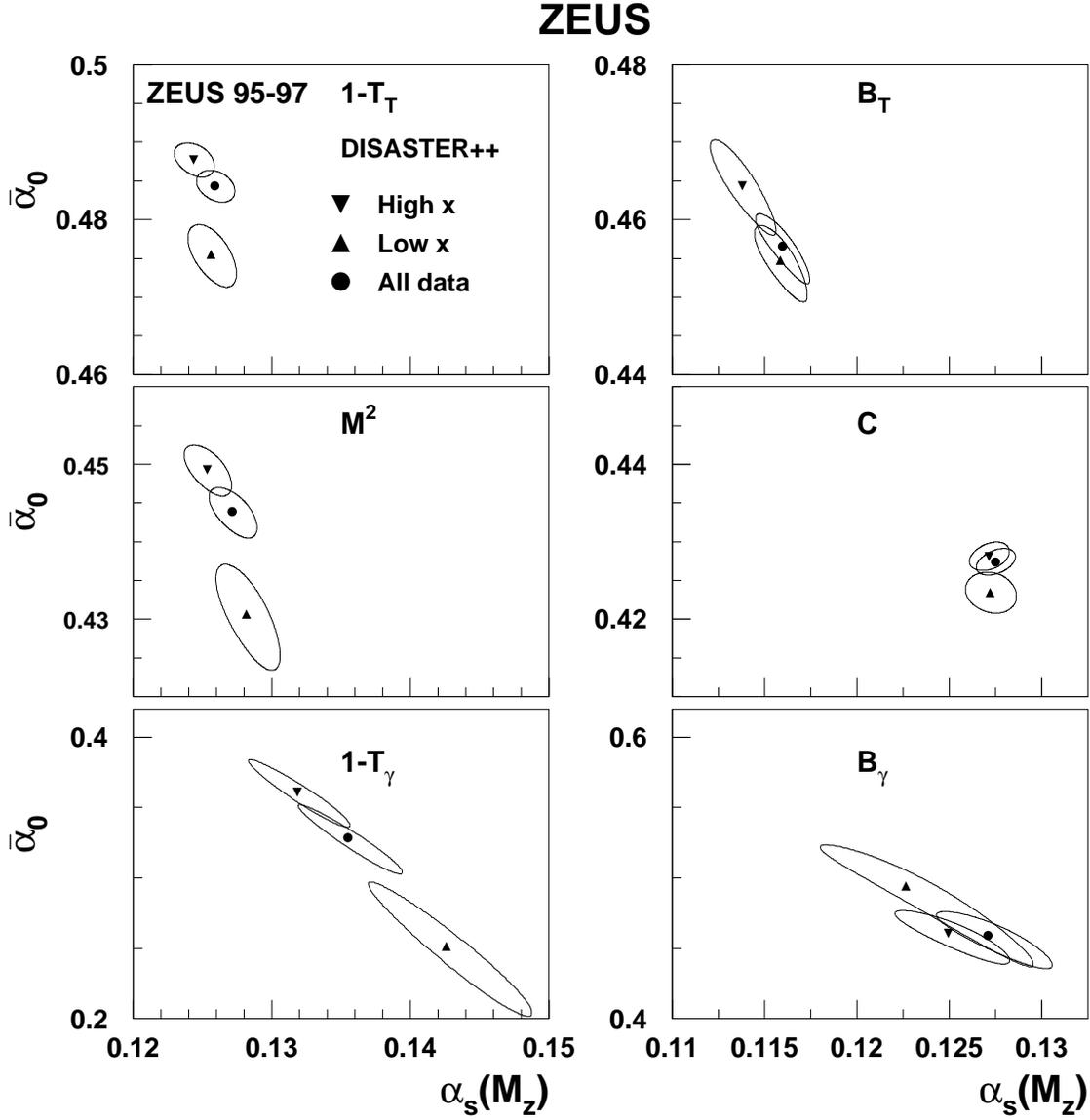}} \caption{Contour
      plots for \Alsmz\ and \albar\ fitted to the mean values of the
      event-shape variables. The fits are based on DISASTER++, with
      $\mathcal{E}_{lim}=0.25 Q$.  The contours show the
      one-standard-deviation limits determined using statistical
      uncertainties only.  The high-$x$ and low-$x$ selections are as defined in 
      Table 1, while `all data' uses  all measured bins for 
      $Q^2 > 80\gev^2$.
      }  \label{fig:4} \end{center} \end{figure}

From the results using DISASTER++, the following features are observed:
 \begin{itemize}
\item $(1-T_\gamma)$ requires a smaller hadronisation correction
than $(1-T_T)$ (Fig.\ \ref{fig:2}), contrary to the theoretical
expectation that the correction should be similar. This is responsible
for the significantly different \albar\ values for the two thrust
variables;
\item $(1-T_\gamma)$ shows a larger $x$-dependence than $(1-T_T)$; the
$M^2$ and $C$ variables, whose definition does not depend on a choice
of axis, show a small $x$ dependence (Fig.\ \ref{fig:2});
\item a residual $x$ dependence in
the fitted \albar\ value obtained from ($1-T_T$), $(1-T_\gamma)$ and $M^2$ but
not from $B_T$, $B_\gamma$ and $C$ can be seen in Fig. \ref{fig:4}.  
However, the \albar\ values are also
consistent with a similar small $x$ dependence in all four variables; 
\item the \Als\ values from all the variables except ($1-T_\gamma$)
show no significant $x$ dependence.
\end{itemize}

There is an inconsistency between the \albar\ values determined from
$(1-T_{\gamma})$ and from the other variables.  As noted earlier,
there is an $x$ dependence in the data from $(1-T_{\gamma})$ that is
not well described by the present model (Fig.~\ref{fig:4}).
Consequently, the fitted values of \Als\ and \albar\ for this variable
are unlikely to be meaningful.  The anomalous value of \Als\ may also
be due to the fact that DISASTER++ does not take full account of
initial-state gluon radiation or other effects related to the target
remnant, which may affect the direction of the current-region system.

For the variables $(1-T_{T})$, $B_{\gamma}$, $M^2$ and $C$, the fitted
\Als\ values are consistent within the statistical uncertainties.  The
variable $B_T$ gives an \Als\ value that differs from the other
determinations, although its \albar\ value agrees within $\pm10\%$
with those from the other four variables.  The inconsistency cannot be
due to experimental or theoretical systematic uncertainties, including
scale uncertainties, since these act in the same direction for all the
variables.  It may be taken to indicate that $B_T$ has a greater
sensitivity to higher-order corrections than the other variables.

A comparison with other measurements is of interest.  With the
exception of $(1-T_\gamma)$, the present results for \Als\ and \albar\
are consistent with those from $e^+e^-$ data within the substantial
theoretical uncertainties.  Using $e^+e^-$ data from a variety of
experiments, Movilla Fern\'andez et al.\ found good agreement between
the means of the event-shape variables as a function of $\sqrt{s}$; in
contrast with the observations of this paper, they obtained \Als\
values that were consistent within statistical errors for all the
variables studied~\cite{delphi,eventshape1,eeshape}.  The \albar\
values were likewise mutually compatible, with the possible exception
of that from $M^2$.  An overall consistency was claimed which
confirmed the validity of the model in the $e^+e^-$ context and thus
enabled an overall experimental value for \Als\ to be given.

In summary, the power correction method applied in DIS gives
consistent values for \Als\ for the event-shape variables $(1-T_T)$,
$B_\gamma$, $M^2$ and $C$.  The \albar\ values for these variables
agree to within $\pm10\%$, which is consistent with the precision
claimed for the model.  The variables $(1-T_\gamma)$ and $B_T$ give,
respectively, \albar\ and \Als\ values that are inconsistent with the
other determinations.   It must be concluded, therefore,
that the power-correction model does not consistently describe all the
shape variables in DIS.  Consequently, no 
average \Als\ or \albar\ values are quoted.

\section{Summary}

A measurement has been made of the mean values of the event-shape
variables thrust $(T)$, broadening $(B)$, normalised jet mass $(M^2)$
and $C$-parameter, using the ZEUS detector at HERA.  The variables $T$
and $B$ were determined relative to the virtual photon axis and the
thrust axis.  The events were analysed in the Breit frame for the
kinematic range $ 6\times 10^{-4} < x < 0.6 $, $ 10 < Q^2 <
20\,480\gev^2$ and $ 0.04 < y < 0.95 $.  The data are successfully
described by the ARIADNE Monte Carlo model.

The $Q$ dependences of the mean event shapes have been fitted to NLO
calculations from perturbative QCD with the DISASTER++ and DISENT
programs together with the Dokshitzer-Webber non-perturbative power
corrections, with the aim of determining \Alsmz\ and $\albarmuI$.
Such a model should give values of \Als\ and \albar\ that are
independent of the shape variable.  Neither DISASTER++ nor DISENT
fulfils these requirements for all variables.

Using DISASTER++, consistent values of \Als\ are obtained for the shape variables $(1-T_T)$, $B_\gamma$, $M^2$ and $C$, with \albar\ values
that agree to within $\pm10\%$.  The \albar\ value from $(1-T_\gamma)$
and the \Als\ value from $B_T$ are in disagreement with the other
determinations.  With the exception of $(1-T_\gamma)$, the present
values are consistent with those measured in $e^+e^-$ annihilation, to
within the theoretical uncertainties.  There is consistency with 
the results from H1.

The power correction method provides a successful description of the
data for all event-shape variables studied.  Nevertheless, the lack of
consistency of the \Als\ and \albar\ determinations obtained in deep
inelastic scattering, together with the dependence of the results on
Bjorken $x$, suggest the importance of higher-order processes that are
not yet included in the model employed in this analysis.  These
effects must be understood before a reliable value of \Als\ can be
quoted using the present method.

\section*{Acknowledgements}
It is once again a pleasure to thank the DESY directorate and staff
for their unfailing support.  The outstanding efforts of the HERA
machine group are likewise gratefully acknowledged, as also are the
many technical contributions from members of the ZEUS institutions who
are not listed as authors.  We are indebted to M. Dasgupta, D.
Graudenz, G. Salam and M. Seymour for many invaluable discussions.

\newpage

\begin{table} \begin{center} 
 \begin{tabular}{||c|c|c||c|c|c||}
   \hline Bin&$Q^{2}\;(\gev^2$)&$x$&Bin&$Q^2\;(\gev^{2}$)&$x$\\
   \hline 1&10 - 20 &~0.0006 - 0.0012&9 & 160 - 320 & 0.0024 - 0.010\\ \hline
   2&10 - 20 &~0.0012 - 0.0024&10 & 160 - 320 & ~0.01 - 0.05 \\ \hline
   3&20 - 40 &~0.0012 - 0.0024&11 & 320 - 640 & ~0.01 - 0.05 \\ \hline
   4&20 - 40 &0.0024 - 0.010  &12 & 640 - 1280 & ~0.01 - 0.05 \\ \hline
   5&40 - 80 &~0.0012 - 0.0024&13 & 1280 - 2560 & ~0.025 - 0.150 \\ \hline
   6&40 - 80 &0.0024 - 0.010  &14 & 2560 - 5120 & ~0.05 - 0.25 \\ \hline
   7& ~80 - 160 &0.0024 - 0.010 &15 & 5120 - 10240 & ~0.06 - 0.40 \\ \hline
   8& ~80 - 160 & ~~~0.01 - 0.050 &16 & 10240 - 20480 & ~0.10 - 0.60 \\ \hline
      \end{tabular} 
      \caption{The kinematic boundaries of the bins in $x$ and $Q^2$.
      The power-correction fits use bins 7 - 16, apart from the fits 
      denoted `high $x$' and `low $x$', which omit bins
      7, 9 and bins 8, 10, respectively.}  \label{table:kinematics} 
      \end{center}
      \end{table}

\newpage
\newcommand{\coltitle}[1]{ \mbox{\small \rule[-1.5ex]{0ex}{5ex} #1 }}
%
%
\begin{sidewaystable}
\begin{center}
{\scriptsize
\begin{tabular}{|c|c|c|c|c|c|c|}
\hline
\coltitle{Bin}& \coltitle{$1-T_T$} & \coltitle{$B_T$} & \coltitle{$M^2$} & 
\coltitle{$C$} & \coltitle{$1-T_\gamma$} & \coltitle{$B_\gamma$} \\

\hline
\hline

 1 & $ 0.1595 \pm0.0010 \pm0.0046 $  & $ 0.2172 \pm0.0009 \pm0.0060 $  & $ 0.0967 \pm0.0006 \pm0.0039 $  & $ 0.540  \pm0.003  \pm0.014  $  & $ 0.508  \pm0.002  \pm0.009  $  & $ 0.3932 \pm0.0010 \pm0.0066 $ \\
 2 & $ 0.1609 \pm0.0010 \pm0.0041 $  & $ 0.2185 \pm0.0009 \pm0.0059 $  & $ 0.0965 \pm0.0006 \pm0.0042 $  & $ 0.544  \pm0.003  \pm0.013  $  & $ 0.499  \pm0.002  \pm0.018  $  & $ 0.3892 \pm0.0010 \pm0.0108 $ \\
 3 & $ 0.1672 \pm0.0011 \pm0.0078 $  & $ 0.2223 \pm0.0010 \pm0.0094 $  & $ 0.0924 \pm0.0007 \pm0.0054 $  & $ 0.565  \pm0.003  \pm0.023  $  & $ 0.458  \pm0.003  \pm0.003  $  & $ 0.3731 \pm0.0011 \pm0.0013 $ \\
 4 & $ 0.1708 \pm0.0009 \pm0.0102 $  & $ 0.2248 \pm0.0008 \pm0.0123 $  & $ 0.0941 \pm0.0005 \pm0.0079 $  & $ 0.575  \pm0.002  \pm0.031  $  & $ 0.437  \pm0.002  \pm0.004  $  & $ 0.3644 \pm0.0009 \pm0.0044 $ \\
 5 & $ 0.1620  \pm0.0020  \pm0.0070  $  & $ 0.2155 \pm0.0014 \pm0.0056 $  & $ 0.0871 \pm0.0009 \pm0.0032 $  & $ 0.556  \pm0.004  \pm0.015  $  & $ 0.426  \pm0.004  \pm0.003  $  & $ 0.3580  \pm0.0020  \pm0.0020  $ \\
 6 & $ 0.1679 \pm0.0010 \pm0.0114 $  & $ 0.2204 \pm0.0009 \pm0.0122 $  & $ 0.0880 \pm0.0006 \pm0.0060 $  & $ 0.572  \pm0.003  \pm0.033  $  & $ 0.3968 \pm0.0024 \pm0.0015 $  & $ 0.3452 \pm0.0011 \pm0.0012 $ \\
\hline
 7 & $ 0.1500 \pm0.0004 \pm0.0050 $  & $ 0.2040 \pm0.0004 \pm0.0049 $  & $ 0.0791 \pm0.0002 \pm0.0030 $  & $ 0.5314 \pm0.0010 \pm0.0143 $  & $ 0.3547 \pm0.0010 \pm0.0031 $  & $ 0.3238 \pm0.0005 \pm0.0016 $ \\
 8 & $ 0.1536 \pm0.0005 \pm0.0110 $  & $ 0.2061 \pm0.0004 \pm0.0097 $  & $ 0.0808 \pm0.0003 \pm0.0084 $  & $ 0.5403 \pm0.0013 \pm0.0300 $  & $ 0.3149 \pm0.0012 \pm0.0190 $  & $ 0.3031 \pm0.0006 \pm0.0096 $ \\
 9 & $ 0.1322 \pm0.0007 \pm0.0013 $  & $ 0.1866 \pm0.0006 \pm0.0020 $  & $ 0.0703 \pm0.0004 \pm0.0008 $  & $ 0.483  \pm0.002  \pm0.006  $  & $ 0.3130  \pm0.0020  \pm0.0050  $  & $ 0.3009 \pm0.0009 \pm0.0029 $ \\
10 & $ 0.1347 \pm0.0006 \pm0.0054 $  & $ 0.1875 \pm0.0005 \pm0.0053 $  & $ 0.0718 \pm0.0003 \pm0.0051 $  & $ 0.489  \pm0.002  \pm0.017  $  & $ 0.2742 \pm0.0013 \pm0.0071 $  & $ 0.2789 \pm0.0007 \pm0.0045 $ \\
11 & $ 0.1150 \pm0.0008 \pm0.0027 $  & $ 0.1676 \pm0.0007 \pm0.0033 $  & $ 0.0628 \pm0.0004 \pm0.0032 $  & $ 0.431  \pm0.002  \pm0.010  $  & $ 0.242  \pm0.002  \pm0.005  $  & $ 0.2574 \pm0.0011 \pm0.0026 $ \\
12 & $ 0.0982 \pm0.0012 \pm0.0017 $  & $ 0.1489 \pm0.0011 \pm0.0024 $  & $ 0.0543 \pm0.0007 \pm0.0017 $  & $ 0.375  \pm0.003  \pm0.006  $  & $ 0.210  \pm0.003  \pm0.004  $  & $ 0.235  \pm0.002  \pm0.002  $ \\
13 & $ 0.086  \pm0.002  \pm0.002  $  & $ 0.133  \pm0.002  \pm0.002  $  & $ 0.0487 \pm0.0010 \pm0.0021 $  & $ 0.330  \pm0.005  \pm0.006  $  & $ 0.168  \pm0.004  \pm0.008  $  & $ 0.203  \pm0.003  \pm0.003  $ \\
14 & $ 0.078  \pm0.003  \pm0.004  $  & $ 0.122  \pm0.003  \pm0.005  $  & $ 0.043  \pm0.002  \pm0.002  $  & $ 0.295  \pm0.009  \pm0.011  $  & $ 0.120  \pm0.007  \pm0.013  $  & $ 0.167  \pm0.005  \pm0.006  $ \\
15 & $ 0.069  \pm0.005  \pm0.004  $  & $ 0.112  \pm0.006  \pm0.004  $  & $ 0.039  \pm0.003  \pm0.003  $  & $ 0.260  \pm0.015  \pm0.012  $  & $ 0.125  \pm0.012  \pm0.015  $  & $ 0.160  \pm0.009  \pm0.012  $ \\
16 & $ 0.062  \pm0.012  \pm0.015  $  & $ 0.104  \pm0.013  \pm0.017  $  & $ 0.030  \pm0.007  \pm0.009  $  & $ 0.233  \pm0.035  \pm0.035  $  & $ 0.058  \pm0.022  \pm0.031  $  & $ 0.110  \pm0.018  \pm0.018  $ \\
\hline
\end{tabular}
} 
%
\caption{Mean event-shape variables in the bins defined in Table \ref{table:kinematics}.
The first uncertainty is statistical and the second is systematic.}
\label{table:tz.means}
\end{center}
\end{sidewaystable}
%
%

\begin{table}
\begin{center}
\begin{tabular}{|c||c|c|c|c|c|c|}
\hline
Variable&$1-T_{T}$&$B_{T}$&$M^2$&$C$&$1-T_{\gamma}$&$B_{\gamma}$\\
\hline\hline
 $\boldsymbol{\alpha_{s}}\mathbf{(M_Z)}$ 
&$\mathbf{ 0.1258} $
&$ \mathbf{0.1159} $
&$ \mathbf{0.1271} $
&$ \mathbf{0.1274}  $  
&$ \mathbf{0.1354} $
&$ \mathbf{0.1270} $\\
\hline
\hline
{\it stat.~error\/}
&$\pm0.0013  $ 
&$\pm0.0013  $ 
&$\pm0.0016  $ 
&$\pm0.0010  $  
&$\pm0.0028  $
&$\pm0.0026  $\\
\hline
{\it stat.$+$sys.~unc.\/}
&$\pm0.0040  $ 
&$\pm0.0026  $ 
&$\pm0.0040  $ 
&$\pm0.0021  $  
&$\pm0.0132  $
&$\pm0.0102  $\\
\hline
$\chi^2 / dof$
&$2.8  $ 
&$1.3  $ 
&$ 2.1 $ 
&$ 1.5 $  
&$ 2.5 $
&$1.8  $\\
\hline {\it correlation } &$ -0.25 $&$ -0.80 $&$ -0.60 $&$ 
        0.26 $&$ -0.93 $&$ -0.84 $\\
\hline
\hline
$ x_{F} = 0.5$
&$-0.0007  $
&$-0.0001  $
&$-0.0004  $
&$-0.0007  $
&$+0.0088 $
&$+0.0025  $\\
\hline
$ x_{F} = 2.0$
&$ +0.0009 $
&$ +0.0008 $
&$ +0.0003 $
&$ +0.0007 $
&$ -0.0008  $
&$ +0.0036 $\\
\hline
$ x_{\mu} = 0.5$
&$-0.0068  $ 
&$-0.0067  $ 
&$-0.0080  $ 
&$-0.0067  $  
&$ -0.0088 $
&$-0.0222  $\\
\hline
$ x_{\mu} = 2.0$
&$ +0.0083 $ 
&$ +0.0081 $ 
&$ +0.0090 $ 
&$ +0.0082 $  
&$ +0.0084  $
&$ +0.0053 $\\
\hline
${\cal{M}} = 1.19$
&$ +0.0025  $
&$ +0.0018 $
&$ +0.0024 $
&$ +0.0029 $  
&$ +0.0032 $
&$ +0.0014 $\\
\hline
$ {\cal{M}} = 1.79$
&$-0.0021  $
&$-0.0017  $
&$-0.0022  $
&$-0.0025  $  
&$-0.0027  $
&$-0.0012  $\\
\hline
$ \mu_{I}=1~\gev$
&$+0.0054  $
&$+0.0042  $
&$+0.0053  $
&$+0.0063  $  
&$+0.0069  $
&$+0.0029  $\\
\hline
$ \mu_{I}=4~\gev$
&$-0.0059  $
&$-0.0047  $
&$-0.0061  $
&$-0.0068  $  
&$-0.0075  $
&$-0.0039  $\\
\hline
\hline
{\it E-scheme}
&$+0.0040  $
&$+0.0028  $
&$+0.0029  $
&$+0.0030  $  
&$+0.0025  $
&$+0.0010  $\\
\hline
\hline
$ \mathbf{Total}$
&$\mathbf{+0.0117}  $
&$\mathbf{+0.0101}  $
&$\mathbf{+0.0118}  $
&$\mathbf{+0.0114}  $
&$\mathbf{+0.0197}  $
&$\mathbf{+0.0128}  $  \\ 
$ \mathbf{uncertainty}$
&$\mathbf{-0.0101}  $
&$\mathbf{-0.0088}  $
&$\mathbf{-0.0111}  $
&$\mathbf{-0.0101}  $
&$\mathbf{-0.0178}  $
&$\mathbf{-0.0248}  $  \\ 
\hline
\end{tabular}
\caption{Fitted results for $\alpha_{s}(M_Z)$ using the NLO prediction
from DISASTER++ and $\mathcal{E}_{lim} = 0.25Q$.  The quoted $\chi^2$
is that from the offset-method fit using statistical uncertainties and
DISASTER++.  The third line is the total experimental \error\ from the
statistical and experimental systematic \errors\ added in quadrature.
The quoted $\chi^2$ is from the offset-method fit using statistical
\errors\ and DISASTER++.  The fifth row gives the correlation
coefficients between the fitted values of $\alpha_{s}(M_Z)$ and \albar\
(see next Table).  The $x_{F}$, $x_{\mu}$, $\cal{M}$, and $\mu_{I}$
rows give the theoretical systematic \errors\ due to variations on the
fragmentation and renormalisation scales, the Milan factor and the
lower limit for the perturbative calculation, respectively; the
$x_{F}$ and $x_{\mu}$ values denote factors by which the respective
scale values are varied.  The systematic effect of using the
$E$-scheme rather than the $P$-scheme is given in the final row of
systematic \errors.  The total \error\ is the total experimental
uncertainty added in quadrature with the theoretical uncertainties. }
\label{table:finalfitnew1}
\end{center}
\end{table}
\begin{table}
\begin{center}
\begin{tabular}{|c||c|c|c|c|c|c|}
\hline
Variable&$1-T_{T}$&$B_{T}$&$M^2$&$C$&$1-T_{\gamma}$&$B_{\gamma}$\\
\hline\hline
$\boldsymbol{\overline{\alpha_{0}}}\;\mathbf{(2~GeV)}$ 
&$\mathbf{0.4843}  $
&$\mathbf{0.4566}  $
&$\mathbf{0.4440}  $
&$\mathbf{0.4274} $  
&$\mathbf{0.3286} $
&$\mathbf{0.4593}  $\\
\hline
\hline
{\it stat.~error\/}
&$\pm0.0020  $ 
&$\pm0.0041  $ 
&$\pm0.0030  $ 
&$\pm0.0017  $  
&$\pm0.0187  $
&$\pm0.0171  $\\
\hline
{\it stat.$+$sys.~error\/}
&$\pm0.0264  $
&$\pm0.0139  $
&$\pm0.0439  $
&$\pm0.0144  $  
&$ \pm0.0993 $
&$\pm0.0815  $\\
\hline
\hline
$ x_{F} = 0.5$
&$+0.0030  $
&$+0.0019  $
&$+0.0006  $
&$+0.0018  $
&$-0.1173  $
&$-0.0491  $\\
\hline
$ x_{F} = 2.0$
&$-0.0017  $
&$-0.0034  $
&$+0.0002  $
&$-0.0006  $
&$+0.0421  $
&$-0.0128  $\\
\hline
$ x_{\mu} = 0.5$
&$+0.007  $
&$+0.056  $
&$+0.016  $
&$+0.007  $  
&$+0.009  $
&$+0.256  $\\
\hline
$ x_{\mu} = 2.0$
&$-0.002  $
&$-0.033  $
&$-0.007  $
&$-0.003  $  
&$+0.029  $
&$-0.135  $\\
\hline
${\cal{M}} = 1.19$
&$+0.0363 $
&$+0.0390 $
&$+0.0223 $
&$+0.0235 $
&$-0.0402  $
&$+0.0250 $\\
\hline
$ {\cal{M}} = 1.79$
&$-0.0282  $
&$-0.0296  $
&$-0.0192  $
&$-0.0200  $  
&$+0.0187  $
&$-0.0197  $\\
\hline
\hline
{\it E-scheme}
&$+0.0163  $
&$+0.0101  $
&$+0.0134  $
&$+0.0127  $  
&$+0.0052  $
&$+0.0103  $\\
\hline
\hline
$ \mathbf{Total}$
&$\mathbf{+0.0483}  $
&$\mathbf{+0.0706}  $
&$\mathbf{+0.0535}  $
&$\mathbf{+0.0312}  $
&$\mathbf{+0.1137}  $
&$\mathbf{+0.2700}  $  \\ 
$\mathbf{uncertainty}$
&$\mathbf{-0.0387}  $
&$\mathbf{-0.0467}  $
&$\mathbf{-0.0484}  $
&$\mathbf{-0.0248}  $
&$\mathbf{-0.1589}  $
&$\mathbf{-0.1664}  $ \\
\hline
\end{tabular}
\rm
\caption{Fitted results for \albar\, defined at $\mu_I = 2$
GeV, using the NLO prediction from DISASTER++ and $\mathcal{E}_{lim} =
0.25Q$.  Other definitions are given in the caption 
to Table \ref{table:finalfitnew1}.}
\label{table:finalfitnew2}
\end{center}
\end{table}

\end{document}